\documentclass[prb,twocolumn,showpacs]{revtex4}
\usepackage{graphicx}

\begin{document}

\title{Effects of pressure on the superconducting properties of magnesium diboride }
\author{X. J. Chen,\thanks{Present address: Department of Physics, Kent State University, 
Kent OH 44242} H. Zhang, and H.-U. Habermeier}
\affiliation{Max-Planck-Institut f\"{u}r Festk\"{o}rperforschung, D-70569 Stuttgart, Germany}
\date{Received 21 May 2001}

\begin{abstract}
We discuss the effects of hydrostatic pressure on the superconducting properties of MgB$_{2}$ 
within the framework of Eliashberg theory. By considering the pressure dependences of all 
parameters appearing in the McMillan formula, we show that the calculated pressure derivative of 
$T_{c}$ as well as the variation of $T_{c}$ with pressure are in good agreement with recent 
measurements. The pressure dependences of the energy gap $\Delta_{0}$, the effective interaction 
strength $N(E_{F})v$, the critical magnetic field $H_{c}(0)$, and the electronic specific heat 
coefficient $\gamma$ are also predicted for this system. A comparison of pressure effect in 
non-transition elements clearly suggests that MgB$_{2}$ is an electron-phonon mediated 
superconductor.
\end{abstract}
\maketitle

\section{INTRODUCTION}

The recent discovery of superconductivity in MgB$_{2}$ (Ref.\ \onlinecite{naga}) has attracted 
considerable interest in the study of this material, both to understand the mechanism of 
superconductivity and to explore other properties of MgB$_{2}$ and related materials. The high 
transition temperature $T_{c}\approx$ 40 K in this material offers other possibility for finding 
high-$T_{c}$ superconductivity in some binary intermetallic compounds besides cuprates and 
$C_{60}$-based compounds. Meanwhile, the high critical currents observed in MgB$_{2}$ thin films 
\cite{eom} and wires\cite{jin} reveal that MgB$_{2}$ belongs to a new class of low-cost, 
high-performance superconducting materials for magnets and electronic applications.

Measurements of the isotope effect and of the influence of pressure on the transition temperature 
and critical field of superconductors yield information on the interaction causing 
superconductivity. Indeed, the pressure (or volume) and the mass number would seem to be the only 
variables whose effect might be capable of immediate theoretical interpretation. By observing 
how pressure changes the parameters of the lattice in the normal state and in the superconducting 
state, and comparing the measurements with the theoretical predictions, one can test the validity 
of some theoretical models. Olsen $et$ $al.$ \cite{olse1} have shown that the volume (V) 
dependence of effective interaction $N(E_{F})v$, $d\ln N(E_{F})v/d\ln V$, can be scaled well with 
the deviation $\xi$ from the full isotope effect where $\xi$ is defined by 
$T_{c}\propto M^{-0.5(1-\xi)}$ in superconducting metals. Bud'ko $et$ $al.$ \cite{bud} and Hinks 
$et$ $al.$ \cite{hink} reported a sizeable isotope effect for B ($\alpha_{B}$=0.26(3) or 0.30(1)) 
in newly discovered superconductor MgB$_{2}$. Although the total isotope coefficient 
$\alpha$=0.32(1) (Ref.\ \onlinecite{hink}) is smaller than the canonical BCS value of 0.5, 
it is the same as that in Cd (Ref.\ \onlinecite{palm}). The isotope effect along with other 
measurements such as inelastic neutron scattering,\cite{sst,osbo} tunneling,\cite{kara} NMR 
(Ref.\ \onlinecite{kote}), and specific heat\cite{krem,walt,bouq} confirmed that MgB$_{2}$ is 
an electron-phonon mediated $s$-wave superconductor.

Soon after the discovery of superconductivity in MgB$_{2}$, the effect of pressure on $T_{c}$ was 
studied by two groups\cite{mont,lore1} by resistivity or $ac$ susceptibility measurements. Both 
groups observed a decrease of $T_{c}$ with increasing pressure, with initial pressure derivative 
$dT_{c}/dP$ of --0.8 K/GPa (Ref.\ \onlinecite{mont}) or --1.6 K/GPa (Ref.\ \onlinecite{lore1}), 
respectively. Moreover, Monteverde $et$ $al.$ \cite{mont} found that the superconductivity is 
not destroyed by applying high pressure up to 25 GPa, at which point $T_{c}$ is as high as 21 K. 
A somewhat larger $dT_{c}/dP$ of --2.0 K/GPa was late reported by Saito $et$ $al.$\cite{sait} from 
high-pressure resistivity measurements. Using a He-gas apparatus, Tomita $et$ $al.$\cite{tomi} 
determined a $dT_{c}/dP$ of --1.11 K/GPa under pure hydrostatic pressure conditions. In order to 
find the reason why the reported values of $dT_{c}/dP$ are different among different groups, 
Lorenz $et$ $al.$ \cite{lore2} carried out high pressure experiments on MgB$_{2}$ samples with 
different $T_{c}$'s at ambient pressure and different pressure media. $T_{c}$ was found to 
decrease linearly over the whole pressure range (0--1 GPa). In the He environment, the two samples
with the initial $T_{c}$=39.2 and 37.4 K yield the pressure derivatives of --1.07 and --1.45 
K/GPa, respectively. The former is obviously very close to that of Tomita $et$ $al.$\cite{tomi} 
The latter approaches their previous data,\cite{lore1} which was obtained by using the Fluorinert 
FC77 as pressure medium. They therefore concluded that the variation in the value of $dT_{c}/dP$ 
by various groups results from the differences in sample preparation conditions. The value of 
$dT_{c}/dP\simeq $--1.1 K/GPa is then confirmed to give the true hydrostatic pressure dependence 
of $T_{c}$ in MgB$_{2}$.

Two theoretical models have been tried to describe the systematics of the behavior of $T_{c}$ 
under pressure in MgB$_{2}$. Based on the theory of hole superconductivity, Hirsch \cite{hirs} 
predicted an increased $T_{c}$ with the decrease of B-B intraplane distance under the application 
of in-plane biaxial pressure. However, this prediction has not been confirmed experimentally yet. 
No uniaxial pressure measurement was reported due to the extreme difficulty in growing MgB$_{2}$ 
single crystal. The experiments of hydrostatic pressure effect on $T_{c}$ do not particularly 
support this theory provided that no charge transfer between the Mg and B layers occurs. 
Alternatively, the experimental results have been analyzed \cite{lore1,tomi,loa,vogt} by using the
McMillan formula \cite{mcmi} derived from Eliashberg theory,\cite{elia} supporting electron-phonon 
mediated superconductivity. Interestingly, Loa and Syassen \cite{loa} analyzed the pressure effect 
on $T_{c}$ from McMillan formula on the basis of their calculated elastic and electronic structure 
data. Assuming that the electron-ion matrix element $I$ is pressure independent, they found that 
the pressure effect on $T_{c}$ is in good agreement with experimental data by using a lattice 
Gr\"{u}neisen parameter $\gamma_{G}$=1. These assumptions, however, deserve some refinements. 
Recent band structure calculations suggest that MgB$_{2}$ is a traditional $sp$ metal 
superconductor.\cite{kort,pick,kong} The pressure dependence of $I$ has long been an interesting 
issue of the research of pressure effect in simple $sp$ metals.\cite{hodd,trof,coul,rott,daco} 
Ziman's calculation of the electron-phonon interaction leads to $<I^{2}>\propto N(E_{F})^{-2}$ at 
least in the limit of long wavelengths.\cite{zima} This then indicated that the consideration of 
the pressure dependence of $I$ would be important for better understanding the superconducting 
properties of MgB$_{2}$ under pressure. On the other hand, it has been found\cite{tomi} 
that the choice of lattice Gr\"{u}neisen parameter $\gamma_{G}$ is crucially important in 
explaining both the magnitude and the sign of the pressure derivative of $T_{c}$ when using 
McMillan formula. The value of $\gamma_{G}$=1 in the calculation of Loa and Syassen is obviously 
lower than those reported recently.\cite{roun,gonc} The pressure dependence of the effective 
electron-electron Coulomb repulsion $\mu^{*}$ appearing in the McMillan formula is usually 
neglected in previous studies due to the assumption of the small change of $\mu^{*}$ compared with 
that of the electron-phonon coupling parameter $\lambda$ (Ref.\ \onlinecite{seid}). However, the 
magnitude of $\mu^{*}$ is also of interest in connection with the possibility that superconductivity 
may be destroyed by pressure.\cite{seid,bo} It was argued that the pressure dependence of $\mu^{*}$ 
makes a significant contribution to the behavior of $T_{c}$ under very high pressures and must 
be handled carefully.\cite{daco,gubs,papa}

In this paper we discuss the pressure dependences of some interested superconducting properties in 
MgB$_{2}$. The outline of this paper is as follows: In Sec. II we presented a theoretical 
approach for pressure effects on the superconducting properties in the simple $sp$ metals 
superconductors. Section III contains the theoretical results obtained and a comparison with 
experiments for MgB$_{2}$. We draw conclusions in Sec. IV.

\section{THEORETICAL FORMULATION}

For our purposes, the relation between $T_{c}$ and microscopic parameters is given adequately by 
the McMillan equation\cite{mcmi}  
\begin{equation}
T_{c}=\frac{\Theta 
_{D}}{1.45}\exp\left[-\frac{1.04(1+\lambda)}{\lambda-\mu^{*}(1+0.62\lambda)}\right]~~, 
\end{equation}
which relates $T_{c}$ to the electron-phonon coupling parameter $\lambda$, the Coulomb repulsion 
strength $\mu^{*}$, and a temperature $\Theta _{D}$ characteristic of the phonons.

Considering the variations of $\Theta _{D}$, $\lambda$, and $\mu^{*}$ with pressure or volume and 
introducing parameters $\varphi = \partial\ln \lambda/\partial\ln V$ and $\phi =\partial\ln 
\mu^{*}/\partial\ln V$, we can get the pressure coefficient of $T_{c}$ 
\begin{eqnarray}
\frac{d\ln T_{c}}{dP}=\frac{\gamma _{G}}{B_{0}} - 
\frac{1.04\lambda(1+0.38\mu^{*})}{[\lambda-\mu^{*}(1+0.62\lambda)]^{2}}\frac{\varphi}{B_{0}} \nonumber \\
+ \frac{1.04\mu^{*}(1+\lambda)(1+0.62\lambda)}{[\lambda-\mu^{*}(1+0.62\lambda)]^{2}}\frac{\phi}{B_{0}} ~~, 
\end{eqnarray} 
where $B_{0}\equiv 1/\kappa _{V}=-\partial P/\partial \ln V$ is the bulk modulus and $\Theta _{D}$ 
is assumed to be proportional to $<\omega^{2}>^{1/2}$ and 
$\gamma _{G}=-\partial\ln <\omega^{2}>^{1/2}/\partial\ln V$ being the effective Gr\"{u}neisen 
parameter.

It is well known that the usual BCS result for the energy gap can be expressed by\cite{bcs}
\begin{equation}
\Delta_{0}=2\Theta _{D}\exp\left[-\frac{1}{N(E_{F})v}\right]~~.
\end{equation} 
Where $N(E_{F})$ is the electronic density of states at the Fermi energy $E_{F}$ and $v$ is the 
pairing potential arising from the electron-phonon interaction. If we renormalize the 
Morel-Anderson result \cite{moan} by introducing the renormalization parameter 
$Z_{n}(0)\equiv 1+\lambda$ into their analysis, the effective interaction strength $N(E_{F})v$ 
can be rewritten as\cite{leav}
\begin{equation}
N(E_{F})v=\frac{\lambda-\mu^{*}}{1+\lambda}~~.
\end{equation}
The logarithmic volume derivative of $N(E_{F})v$ is then given by
\begin{equation}
\frac{d\ln N(E_{F})v}{d\ln V}=\frac{\lambda(1+\mu^{*})}{(\lambda-\mu^{*})(1+\lambda)}\varphi
-\frac{\mu^{*}}{\lambda-\mu^{*}}\phi~~.
\end{equation}
Considering the experimental observations of the pressure dependence of the energy gap of 
superconductor,\cite{jpf} we differentiate Eq. (3) with respect to pressure
\begin{equation}
\frac{d \ln \Delta_{0}}{dP}=\frac{\gamma _{G}}{B_{0}}-\frac{1}{B_{0}}\left[
\frac{\lambda(1+\mu^{*})}{(\lambda-\mu^{*})^{2}}\varphi
-\frac{\mu^{*}(1+\lambda)}{(\lambda-\mu^{*})^{2}}\phi \right]~~.
\end{equation}

BCS expression for the critical field $H_{c}$ at absolute zero temperature is\cite{bcs}
\begin{equation}
\frac{H_{c}(0)^{2}}{8\pi}=2N(E_{F})\omega^{2}\exp\left[-\frac{2}{N(E_{F})v}\right]~~.
\end{equation}
Differentiating Eq. (7) with respect to the pressure, one obtains an expression of the pressure
coefficient of $H_{c}(0)$,  
\begin{equation}
\frac{d\ln H_{c}(0)}{dP}=\frac{d\ln \Delta_{0}}{dP}-\frac{\gamma_{N}}{2B_{0}}+\frac{1}{2B_{0}}~~,
\end{equation}
where $\gamma _{N}=\partial \ln N(E_{F})/\partial \ln V$.

The expressions for $\gamma _{G}$, $\varphi$, and $\phi$ can be integrated to give
\begin{eqnarray}
\Theta _{D}(V) &=& \Theta _{D}(0)\left[V/V_{0}\right]^{-\gamma _{G}}\\ \nonumber
\lambda(V) &=& \lambda(0)\left[V/V_{0}\right]^{\varphi}\\ \nonumber
\mu^{*}(V) &=& \mu^{*}(0)\left[V/V_{0}\right]^{\phi} ~~.
\end{eqnarray}
Here $V$ and $V_{0}$ are the unit cell volumes under the applied pressure and at ambient 
pressure, respectively. These two volumes can be related according to the first-order Murnaghan 
equation of state $V(P)=V(0)(1+B_{0}^{\prime}P/B_{0})^{-1/B_{0}^{\prime}}$. The Eq. (9) is then 
rewritten as
\begin{eqnarray}
\Theta _{D}(P) &=& \Theta _{D}(0)\left[1+\frac{B_{0}^{\prime}P}{B_{0}}\right]^{\gamma 
_{G}/B_{0}^{\prime}}\\ \nonumber
\lambda(P) &=& \lambda(0)\left[1+\frac{B_{0}^{\prime}P}{B_{0}}\right]^{-\varphi/B_{0}^{\prime}}\\ 
\nonumber
\mu^{*}(P) &=& \mu^{*}(0)\left[1+\frac{B_{0}^{\prime}P}{B_{0}}\right]^{-\phi/B_{0}^{\prime}} ~~.
\end{eqnarray}

From Eqs. (1) and (10) we arrive at the expression for the pressure dependence of $T_{c}$
\begin{equation}
T_{c}(P)=T_{c}\left[\Theta _{D}(P), \lambda(P), \mu^{*}(P)\right]~~.
\end{equation}

Knowing $B_{0}$, $B_{0}^{\prime}$, $\gamma _{G}$, $\gamma _{N}$, $\phi$, and $\varphi$, one can 
evaluate the pressure effects on the superconducting properties, especially the behavior of 
$T_{c}$ under pressure. $B_{0}$ and $B_{0}^{\prime}$ can be obtained from the compressibility data 
determined by neutron or synchrotron x-ray diffractions. A direct experimental determination of 
$\gamma _{G}$ can be made by measuring electron tunneling \cite{keel,yama,zava} or inelastic 
neutron scattering\cite{lech} under high pressure. In general, for metals in which different 
techniques yield similar Gr\"{u}neisen constants, a good approximation to $\gamma _{G}$ is 
provided by the room temperature value determined from the Gr\"{u}neisen equation 
\begin{equation}
\gamma _{G}=\frac{\alpha _{V}V_{m}}{\kappa _{V}C_{p}}~~,
\end{equation}    
where $\alpha _{V}$ is the volume coefficient of thermal expansion, $V_{m}$ is the molar volume, 
and $C_{p}$ is the molar heat capacity at constant pressure. The approximation for $\gamma_{G} $ 
of Slater is derived from the pressure derivative of the bulk modulus 
\cite{slat}
\begin{equation} 
\gamma_{G} ^{S} \equiv \frac{B_{0}^{\prime}}{2}-\frac{1}{6}=-\frac{2}{3}
-\frac{1}{2}\frac{V\partial ^{2}P/\partial V^{2}}{\partial P/\partial V}~~.
\end{equation}

The formula for $\mu^{*}$ due to Morel and Anderson \cite{moan} used here is
\begin{equation}
\mu^{*}=\frac{\mu}{1+\mu \ln (E_{F}/\omega_{ph})}~~,
\end{equation} 
with $\mu=0.5\ln[(1+a^{2})/a^{2}]$ and $a^{2}=\pi e^{2}N(E_{F})/k_{F}^{2}$, from which we 
evaluate the volume dependence of $\mu^{*}$ as
\begin{equation}
\phi=\mu^{*}\left[\frac{2}{3}-\gamma_{G}-\frac{1-e^{-2\mu}}{2\mu^{2}}(\gamma_{N}+\frac{2}{3})\right]~~.
\end{equation}
Here the variation of $k_{F}$ with volume has been calculated from the fundamental definition 
$k_{F}=(3\pi^{2}Z/V)^{1/3}$ with $Z$ the valency. Unfortunately, to the best of our knowledge 
$\gamma_{N}$ has never measured directly for any superconductor in the case of free electron gas 
it would have a value of $2/3$.  Using the expression given by Migdal \cite{migdal} for the 
electronic specific heat coefficient $\gamma$ one obtains for the electronic Gr\"{u}neisen 
parameter
\begin{equation}
\gamma_{e}=\frac{\partial \ln \gamma}{\partial \ln V}=\gamma_{N}+\frac{\lambda}{1+\lambda}\varphi~~.
\end{equation} 
The electronic Gr\"{u}neisen parameter $\gamma _{e}$ is usually deduced from measurements through 
the simple relation \cite{barr}
\begin{equation}
\gamma _{e}=\frac{\alpha _{e}V_{m}}{\kappa _{V}C_{e}}~~.
\end{equation}
Here $\alpha _{e}$ is the contribution to the expansion coefficient from the electrons at lower 
temperatures and $C_{e}$ is the electronic heat capacity. A theoretical estimate of $\gamma _{e}$ 
can also be given from the measurement of the volume dependence of the orbital susceptibility 
\cite{fawc} or from band-structure considerations.\cite{flet}

The electron-phonon coupling parameter $\lambda$ can be expressed as
\begin{equation}
\lambda =\frac{N(E_{F})<I^{2}>}{M<\omega^{2}>}\equiv \frac{\eta}{M<\omega^{2}>}~~,
\end{equation}
where $<I^{2}>$ is the mean-square electron-ion matrix element and $M$ the ionic mass. The 
McMillan-Hopfield parameter $\eta$ (or $N(E_{F})<I^{2}>$) has been regarded as a local 
``chemical'' property of an atom in a crystal. Allen and Dynes \cite{alle} pointed out that 
$\eta $ is the most significant single parameter in understanding the origin of the high 
$T_{c}$ of conventional superconductors. For strong coupling systems, variation in $\eta$ is 
more important than variation of $<\omega^{2}>$ in causing $T_{c}$ to change. Softening 
of $<\omega^{2}>$ often does enhance $T_{c}$, but very high $T_{c}$ should be caused more by 
large $\eta$ than by small $<\omega^{2}>$.

The logarithmic volume derivative of $\lambda$, $\varphi$, is then obtained
\begin{equation}
\varphi=\frac{\partial \ln \eta}{\partial \ln V} + 2\gamma _{G} \equiv S+2\gamma _{G}~~.
\end{equation}
In order to understand how the electronic contribution $\eta=N(E_{F})<I^{2}>$ varies with volume, 
we use the Gaspari-Gyoriffy theory \cite{gasp} for $\eta$, i.e.,
\begin{equation}
\eta=\frac{k_{F}^{2}}{\pi^{2}N(E_{F})}\sum_{l}\frac{2(l+1)\sin^{2}(\delta_{l+1}-
\delta_{l})N_{l}N_{l+1}}{N_{l}^{1}N_{l+1}^{1}}~~,
\end{equation}
where $N_{l}$ is the $l$th angular momentum component of the density of states, $N_{l}^{1}$ is the
$l$th component of the single scatterer density of states evaluated at $E_{F}$, and $\delta_{l}$ 
the phase shift.

For simple metals, the scatterers are assumed to be weak. We can take $N_{l}=N_{l}^{1}$ and 
approximate $\sin^{2}(\delta_{l+1}-\delta_{l})$ by $(\delta_{l+1}-\delta_{l})^{2}$. Eq. (20) is 
rewritten as
\begin{equation}
\eta=\frac{k_{F}^{2}}{\pi^{2}N(E_{F})}\sum_{l}2(l+1)(\delta_{l+1}-\delta_{l})^{2}~~.
\end{equation}
This expression is identical to the pseudopotential formula of McMillan.\cite{mcmi,evan} Assuming 
that the phase shift $\delta_{l}$ does not vary very much under pressure for simple $sp$ 
superconductors, we then obtain
\begin{equation} 
S=-\gamma_{N}-\frac{2}{3}~~.
\end{equation}
The form in Eq. (22) is the same as that of Baryakhtar and Makarov,\cite{bary} who used the 
constant of the electron-phonon interaction of Fr\"{o}hlich and Mitra.\cite{floh} The expression 
is an improvement over the expressions of $S$=0 and $S=-$4/3 obtained by Olsen $et$ $al.$ 
\cite{olsen1} and Seiden,\cite{seid} respectively. It is interesting to notice that 
substitution of $\gamma _{N}$=2/3 into Eq. (22) yields $S=-4/3$. Eq. (22) reduces to the 
expression of Seiden,\cite{seid} who modified McMillan's expression for $\lambda$ somewhat by 
considering the effects of a real lattice spectrum as opposed to the Jellium model. Since the 
electronic Gr\"{u}neisen parameter $\gamma _{e}$ usually varies among different metals even in 
the simple non-transition elements,\cite{palm,barr,flet} we believe that Eq. (22) should 
provide a more reasonable value of $S$ compared with Seiden's formula.

\section{RESULTS AND DISCUSSION}

Using the experimental value of $T_{c}$=39.25 K (Refs.\ \onlinecite{tomi,lore2}), and the 
theoretical estimates of $\lambda$=0.87 and $\mu^{*}$=0.10 (Ref.\ \onlinecite{kong}), we got 
$\Theta_{D}$=860 K from Eq. (1) for MgB$_{2}$. We believe that all these parameters, which will 
enter our calculations, are reliable. For example, the inelastic neutron scattering measurements 
\cite{osbo} provide an estimate of $\lambda\sim $0.9, which is close to that we used. The 
calculated value of $\Theta_{D}$=860 K is in the range from 746 to 1050 K determined from the 
specific heat measurements.\cite{krem,walt,bouq}

We took the structural parameters $B_{0}$=147.2 GPa and $B_{0}^{\prime}$=4 from the measurements 
under the pure hydrostatic pressures up to 0.62 GPa (Ref.\ \onlinecite{jorg}) and under high 
pressures up to 15 GPa (Refs.\ \onlinecite{vogt,gonc}), respectively. To our knowledge no 
inelastic neutron scattering or tunneling data exist for MgB$_{2}$ under hydrostatic pressure. 
We have to use Eq. (12) or (13) for estimating the lattice Gr\"{u}neisen parameter $\gamma_{G}$. 
The measurements of heat capacity\cite{swif} give a $C_{p}$ of 47.80 J/(K mol) at $T$=298.16 K. 
$V_{m}$=1.75$\times 10^{-5}$ m$^{3}$/mol, $\kappa _{V}$=6.79$\times 10^{-12}$ Pa$^{-1}$, and 
$\alpha _{V}$=2.22$\times 10^{-5}$ K$^{-1}$ can be drawn from the neutron diffraction data. 
\cite{jorg} We therefore obtained $\gamma_{G}$=1.2 by using Eq. (12). Based on the first-order 
Murnaghan equation for $V(P)$ and the Slater expression of Eq. (13), we got a somewhat larger 
$\gamma_{G}$ of 1.83 compared to that from Eq. (12). For most simple metals, there is no much 
difference between the room temperature lattice Gr\"{u}neisen parameter given through Eq. (12) 
and the Slater relation.\cite{seid,gsch} It was found that the Slater expression usually can 
yield the reasonable values of $\gamma_{G}$ for most metals.\cite{moru} The only uncertainty 
entering Eq. (12) in our calculation comes from the indirect measurements of the linear 
coefficients of thermal expansion.\cite{jorg} Roundy $et$ $al.$ \cite{roun} reported a value of 
$\gamma_{G}\approx$2.3 from $ab$ $initio$ calculations, which is close to our calculated 
$\gamma_{G}^{S}$ according to Eq. (13). Meanwhile, Goncharov $et$ $al.$ \cite{gonc} 
determined a large $E_{2g}$ mode Gr\"{u}neisen parameter of 2.9$\pm$0.3 from the  measurements 
of Raman spectra under pressure. This value is obviously larger than those derived from Eqs. 
(12) and (13).

\begin{figure}[tbp]
\vspace{-1.0cm}
\begin{center}
\includegraphics[height=9.0cm,width=7.5cm,angle=270]{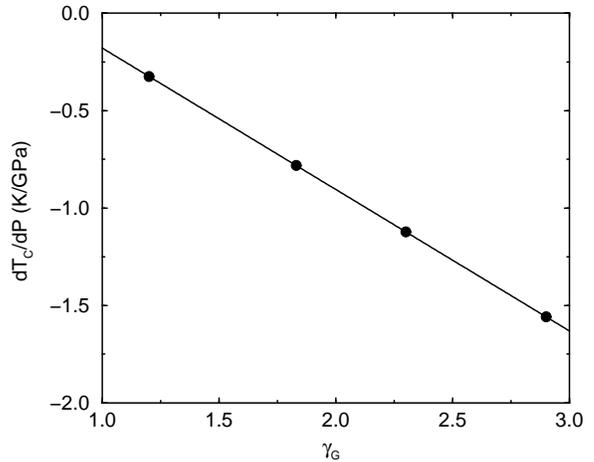}
\end{center}
\caption{ Pressure derivative of $T_{c}$ as a function of the lattice Gr\"{u}neisen parameter 
$\gamma_{G}$ in MgB$_{2}$. The circles show the calculation from the four different values of 
$\gamma_{G}$=1.2, 1.83, 2.3, and 2.9. }
\end{figure}

In the calculations of electronic density of states, Loa and Syassan\cite{loa} found that 
$N(E_{F})$ decreases with pressure at a rate of $d\ln N(E_{F})/dP=$--3.1$\times$10$^{-3}$ 
GPa$^{-1}$. Combining this calculated value and experimental value of $B_{0}$, we got 
$\gamma_{N}$=0.46. The volume dependence of $\mu^{*}$ is then derived from (15) once having 
the values of $\gamma_{G}$ and $\mu$. For simple $sp$ metals, $a^{2}$ has a typical value of 
0.4 (Ref.\ \onlinecite{moan}), which yields $\mu=$0.63. The volume dependence of $\lambda$, 
$\varphi$, is therefore determined from Eqs. (19) and (22). Using $\gamma_{N}$=0.46, we have 
$S$=--1.13 for MgB$_{2}$, which is smaller in magnitude than --4/3 in Seiden's formula for 
simple metals.\cite{seid} For the transition metals, Hopfield \cite{hopf} commented that $S$ 
is a relatively constant quantity with a value of about --3.5. The values of 
$S$=--3.5$\sim$--3.1 obtained by inverting the measured $dT_{c}/dP$ for 
YNi$_{2-x}$M$_{x}$B$_{2}$C (M=Co and Cu) (Ref.\ \onlinecite{loon}) are comparable to that of 
the transition metals, but are larger in magnitude compared to that of MgB$_{2}$.

With the parameters determined above, we have calculated the pressure derivatives of $T_{c}$ for 
MgB$_{2}$ by using Eq. (2). In Fig. 1 we plotted $dT_{c}/dP$ as a function of $\gamma_{G}$ in 
the interested range. It is interesting to note that $\gamma_{G}$ plays a predominant role 
for the pressure effect of $T_{c}$. For the four different $\gamma_{G}$'s considered here 
$dT_{c}/dP$ are negative. The values obtained from $\gamma_{G}$=1.83 and 2.3 are --0.78 and 
--1.12 K/GPa, respectively. These are close to the hydrostatic pressure value of --1.1 K/GPa 
(Refs.\ \onlinecite{tomi,lore2}). Thus the hydrostatic pressure results can be reproduced in 
terms of our present model by using the values of $\gamma_{G}$ obtained from either the Slater 
relation or $ab$ $initio$ calculation. It is difficult to obtain the measured results by using 
$\gamma_{G}$=1 as suggested by Loa and Syassen.\cite{loa} We noticed that a $\gamma_{G}$ of 
2.27 is necessary so as to account for the pressure effect on $T_{c}$ for MgB$_{2}$. As 
emphasized above, all quantities entering Eq. (12) are experimental values and only $\alpha _{V}$ 
was taken from indirect measurements. Thus it is highly expected to operate the thermal expansion 
measurement to yield a direct $\alpha _{V}$. The present results indicate that the range from 
$\gamma_{G}$=1.83 to 2.3 should cover the reasonable choices for the lattice Gr\"{u}neisen 
parameters.

To verify these results, and also to study the behavior of $T_{c}$ as a function of pressure, we 
have performed explicit calculation based on Eq. (11). The theoretical results in the pressure 
range from 0 to 1.0 GPa are shown in Fig. 2. The experimental data points of Tomita $et$ 
$al.$ \cite{tomi} and Lorenz $et$ $al.$ \cite{lore2} measured under hydrostatic pressure 
conditions are also plotted for comparison. It is clearly seen that our calculations agree 
well with the experiments.

\begin{figure}[tbp]
\vspace{-1.5cm}
\begin{center}
\includegraphics[height=9.0cm,width=7.5cm,angle=270]{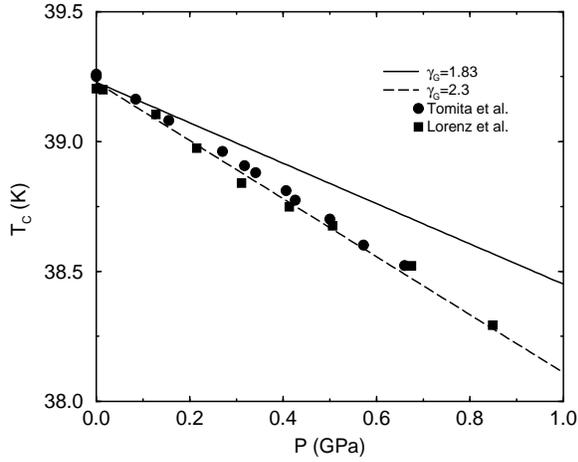}
\end{center}
\caption{ Variation of $T_{c}$ with pressure in the region of 0 to 1.0 GPa of MgB$_{2}$ for 
$\gamma_{G}$=1.83 and 2.3, respectively. The circles and squares represent the hydrostatic 
pressure experimental data taken from the works of Tomita $et$ $al.$ \cite{tomi} and Lorenz $et$ 
$al.$,\cite{lore2} respectively. }
\end{figure}

In Fig. 3 we presented the calculated results as well as the experimental data points of 
Monteverde $et$ $al.$ \cite{mont} and Deemyad $et$ $al.$\cite{deem} measured in the relatively 
high pressure region. Here we assume that phase transitions of all kinds do not occur under 
pressure range that we consider. We noticed that the experimental data points of Deemyad $et$ 
$al.$\cite{deem} and the sample 4 of Monteverde $et$ $al.$\cite{mont} are situated well between 
the two theoretical curves calculated by using $\gamma_{G}$=1.83 and 2.3, respectively. 
Interestingly, the agreement between our theoretical curve calculated by using $\gamma_{G}$=1.83 
and the experimental data points of other samples of Monteverde $et$ $al.$\cite{mont} is seen to 
be reasonable, although there are some scatters among different samples and the reason is not 
clear. Furthermore, although the pressure measurements are limited to the region below 33 GPa, 
it is seen from the inset of Fig. 3 that Eq. (11) continues to describe the pressure dependence 
of $T_{c}$ as high as 100 GPa. Even at this point, the superconductivity is not destroyed by 
pressure in newly discovered superconductor MgB$_{2}$. There was a discrepancy on whether pressure 
can destroy superconductivity.\cite{seid,bo,olsen2,smithc} However, our results support the 
conclusion of Olsen and collaborators\cite{bo,olsen2} that the possibility of destruction of 
superconductivity by the application of sufficiently high pressure most likely does not exist. 
It follows from the comparison of Figs. 2 and 3 that the the pressure effect on $T_{c}$ indeed 
provides a support to the electron-phonon mediated superconductivity in MgB$_{2}$.

\begin{figure}[tbp]
\begin{center}
\includegraphics[width=8.2cm,height=6.5cm,angle=0]{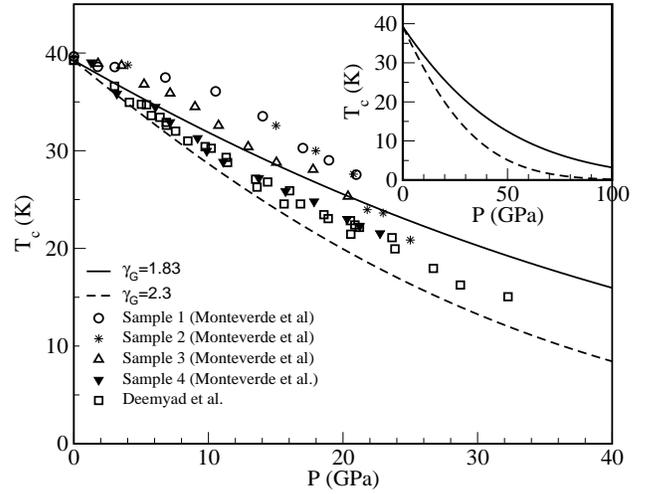}
\end{center}
\caption{ Pressure dependence of the transition temperature in MgB$_{2}$ up to 40 GPa. Experimental 
data are from the works of Monteverde $et$ $al.$\cite{mont} and Deemyad $et$ $al.$\cite{deem} The inset 
is a calculation of $T_{c}$ under pressure up to 100 GPa. }
\end{figure}

In Fig. 4 we presented the normalized $\lambda$ and $\mu^{*}$ as a function of pressure up to 
30 GPa, calculated from Eq. (10) by using $\gamma_{G}$=2.3. The Coulomb pseudopotential
$\mu^{*}$ increases slightly with pressure. Whereas $\lambda$ changes significantly with 
pressure. The contribution from $\mu^{*}$(P) to the variation of $T_{c}$ with pressure is 
much less important than that of $\lambda$(P). Thus in the range from 0 to 30 GPa the 
pressure effect of $T_{c}$ for MgB$_{2}$ is dominated by the competition of $\lambda$ and 
$\Theta _{D}$ (or $<\omega^{2}>^{1/2}$).

\begin{figure}[tbp]
\vspace{-1.0cm}
\begin{center}
\includegraphics[height=9.0cm,width=7.5cm,angle=270]{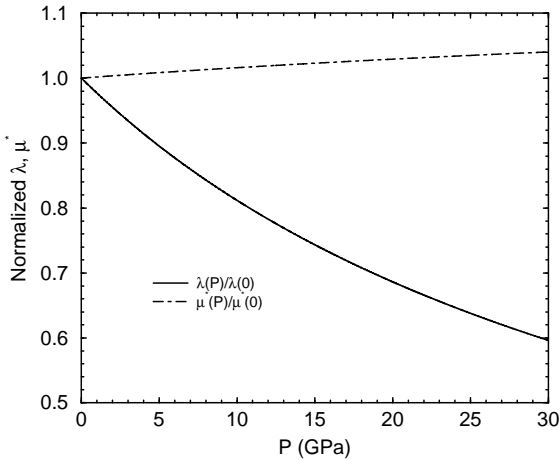}
\end{center}
\caption{ Pressure dependence of normalized Coulomb ($\mu^{*}$) and electron-phonon coupling 
($\lambda$) coupling strengths in MgB$_{2}$ calculated by using $\gamma_{G}$=2.3. }
\end{figure}

\begin{table}[b]
\caption{Pressure dependences of superconducting state parameters in MgB$_{2}$. The units of 
$d\ln X/dP$ (X=$T_{c}$, $\Delta_{0}$, are $H_{c}(0)$) are in 10$^{-2}$ GPa$^{-1}$. }
\label{table. 1}
\begin{ruledtabular}
\begin{tabular}{cccccc}
$\gamma_{G}$ & $\gamma_{e}$ & $\frac{d\ln N(E_{F})v}{d\ln V}$ & $\frac{d\ln
T_{c}}{dP}$ & $ \frac{d \ln \Delta_{0}}{dP}$ & $\frac{d\ln H_{c}(0)}{dP}$ \\
\hline
1.83 & 1.64 & 1.71 & -1.99 & -1.58 & -1.39\\
2.3 & 2.07 & 2.34  & -2.86 & -2.30 & -2.11\\
\end{tabular}
\end{ruledtabular}
\end{table}

Table I contains the calculated values of pressure dependences of superconducting parameters for 
MgB$_{2}$ from Eqs. (5), (6), (8) and (16) by using $\gamma_{G}=1.83$ and $\gamma_{G}=2.3$, 
respectively. The reliable values of $\gamma_{e}$ is readily determined using Eqs. (16), (19), 
and (22). We obtained $\gamma_{e}$=1.64 and 2.07 for MgB$_{2}$, which are close to 
$\gamma_{e}$=1.7 for Pb and for Sn, $\gamma_{e}$=2.0 for Al (Ref.\ \onlinecite{smithc}). The 
negative sign for $d\ln H_{c}(0)/dP$ predicted for MgB$_{2}$ is in agreement with the 
measurements for all simple elements superconductors with the exception of thallium.\cite{levyo} 
For the simple $sp$ metals superconductors, Rohrer\cite{rohrer} has demonstrated that 
$d\ln N(E_{F})v/d\ln V$ must have approximately a value of 2.0. However, it was realized 
\cite{levyo} that the transition metals fail to show such simple behavior. Our estimated 
$d\ln N(E_{F})v/d\ln V$=1.71 and 2.34 for MgB$_{2}$ are comparable to those obtained for simple 
$sp$ metals superconductors.\cite{levyo,rohrer} Early measurements for most simple metals 
\cite{trof,jpf,zava} show that there is difference between the quantities in $d\ln \Delta_{0}/dP$ 
and $d\ln T_{c}/dP$. This can be understood with the aid of the results by Ge\v{i}likman and 
Kresin,\cite{beili} that is, 
$2\Delta_{0}/k_{B}T_{c}=3.52[1+5.3(T_{c}/\omega_{ph})^{2}\ln\omega_{ph}/T_{c}]$. The calculated 
data of MgB$_{2}$ listed in Table I make it possible to support this theory. Since the phonon 
spectrum shifts under pressure, it follows that for all superconductors with 
$2\Delta_{0}/k_{B}T_{c}>3.52$ a change of $2\Delta_{0}/k_{B}T_{c}$ under pressure can be 
expected. It is interesting from the viewpoint of experiment to investigate the tunnel 
characteristics of MgB$_{2}$ under hydrostatic pressure.

\begin{table}[b]
\caption{Experimental values of $d\ln N(E_{F})v/d\ln V$, isotope effect exponent $\alpha$ and 
its derivative $\xi=1-2\alpha$ in nine simple $sp$ metals superconductors. }
\label{table. 2}
\begin{ruledtabular}
\begin{tabular}{ccccc}
Element & Z & $\frac{d\ln N(E_{F})v}{d\ln V}$ & $\alpha$ & $\xi$ \\
\hline
Zn & 2 & 2.0 & 0.37 & 0.26 \\
Cd & 2 & 2.9 & 0.32 & 0.36 \\
Hg($\alpha$) & 2 & 1.7 & 0.50 & 0 \\
Al & 3 & 3.4 & 0.325 & 0.35 \\
Ga & 3 & 1.8 & 0.41 & 0.18 \\
In & 3 & 2.3 & 0.466 & 0.068 \\
Tl & 3 & 0 & 0.49 & 0.02 \\
Sn & 4 & 2.3 & 0.47 & 0.06 \\
Pb & 4 & 2.1 & 0.478 & 0.044\\
\end{tabular}
\end{ruledtabular}
\end{table}

\begin{figure}[tbp]
\vspace{-1.0cm}
\begin{center}
\includegraphics[height=9.0cm,width=7.5cm,angle=270]{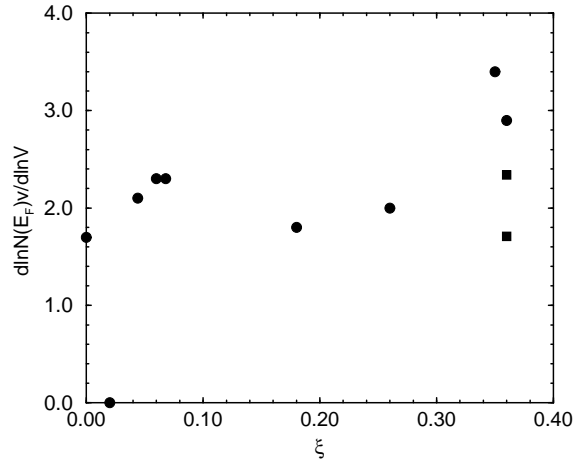}
\end{center}
\caption{Relation between the logarithmic volume derivative of $N(E_{F})v$ and the deviation 
$\xi$ from the full isotope effect exponent $\alpha=0.5(1-\xi)$ in nine simple $sp$ metals 
superconductors and MgB$_{2}$. The two squares are the values for MgB$_{2}$. }
\end{figure}

Figure 5 is a plot of $d\ln N(E_{F})v/d\ln V$ versus the deviation $\xi$ from the full isotope 
effect for nine simple $sp$ metals superconductors as well as MgB$_{2}$. The experimental values 
of $d\ln N(E_{F})v/d\ln V$ for simple metals are chosen from the work of Olsen, Andres, and 
Geballe.\cite{olsen1} The experimental results for isotope effect exponent $\alpha$ and its 
deviation $\xi$ are taken from the works in Refs. [\ \onlinecite{palm,maxw,hake,fass1,fass2,meser}]. 
There are no experimental data available for Al and In now, we took the calculated results from 
Leavens and Carbotte.\cite{leav} We summarized these results for simple $sp$ metals in Table II. 
Olsen $et$ $al.$ \cite{olse1} suggested that $d\ln N(E_{F})v/d\ln V$ is related to the isotope 
effect exponent $\alpha$ in metals superconductors. As seen from Fig. 5 the relation between 
$d\ln N(E_{F})v/d\ln V$ and $\xi$ is not very clear when more data included. An isotope effect 
$\alpha$=0.32(1) (Ref.\ \onlinecite{hink}) in MgB$_{2}$ is consistent with what appears to 
be a systematic variation of $\alpha$ across the non-transition elements. It is well known that 
deviations of the isotope effect exponent from $1/2$ are a measure of the relative strengths 
of the Coulomb and phonon-mediated electron-electron interactions. It is indicated, from the 
compared values of $\xi$ for MgB$_{2}$ (Ref.\ \onlinecite{hink}) with Zn (Ref.\ 
\onlinecite{fass1}), Cd (Ref.\ \onlinecite{palm}), and Al (Ref.\ \onlinecite{leav}), that 
MgB$_{2}$ should be a medium coupling superconductor.

\section{SUMMARY AND CONCLUSIONS}

The major conclusions given by present investigation can be summarized as follows:

(i) A simple expression was derived for the pressure dependences of superconducting properties in 
simple $sp$ superconductor on the basis of McMillan equation. The logarithmic volume derivatives 
of $\lambda$, $\mu^{*}$, and $\Theta _{D}$ can be self-consistently determined from experiments 
and theories. We gave an expression for $\varphi$ from the theory of Gaspari and Gyorffy. 
\cite{gasp} The theory of Morel and Anderson \cite{moan} was used to obtain $\phi$, which makes 
it possible to investigate the pressure dependence of $\mu^{*}$. Neglecting the pressure 
dependence of $\mu^{*}$, the present theoretical model can be reduced to the two popular models 
of Seiden \cite{seid} and Baryakhtar and Makarov \cite{bary} when taking $\gamma_{N}=$ 2/3 and 
neglecting the direct electron-electron interaction, respectively. Furthermore, we obtained an 
explicit expression for the change of $T_{c}$ as a function of pressure with the help of 
Murnaghan equation. The present model enables us to study the pressure behaviors of some 
interested superconducting parameters such as the zero temperature energy gap $\Delta_{0}$, the 
critical field at absolute zero temperature $H_{c}(0)$, the effective interaction strength 
$N(E_{F})v$, and the electronic specific heat coefficient $\gamma$.

(ii) We investigated the pressure effects on superconducting properties in the newly discovered 
superconductor MgB$_{2}$ using our simple approach. It was found that the hydrostatic pressure 
derivative of $T_{c}$ can be reproduced by using the values of $\gamma_{G}$ obtained from 
either the Slater relation or $ab$ $initio$ calculation. The calculated 
$d\ln N(E_{F})v/d\ln V\approx$2.0 in MgB$_{2}$ is close to those obtained in simple $sp$ 
superconductors. The quantitative agreement for the variation of $T_{c}$ with pressure in the 
low pressure region as well as high pressure region is very good when comparing our theoretical 
results with experimental data measured by three groups. The predicted values of 
$d\ln H_{c}(0)/dP$, $d\ln \Delta_{0}/dP$, and $\gamma_{e}$ are also comparable to those in simple 
$sp$ metals superconductors. All these characteristic pressure behaviors allow us to conclude 
that MgB$_{2}$ should be a simple electron-phonon mediated $sp$ superconductor and the mechanism 
in simple $sp$ metals superconductors is also responsible for the superconductivity in MgB$_{2}$. 

\begin{acknowledgments}
The authors acknowledge useful discussions with O. K. Andersen, O. Jepsen, Y. Kong, R. K. 
Kremer, and K. Syassen. We are indebted to J. S. Schilling and J. D. Jorgensen for allowing us 
to use their experimental data prior to publication as well as their valuable comments on the 
manuscript. XJC thanks the MPG for financial support.
\end{acknowledgments}

{\it Note added.--} After submission of this manuscript, the authors have learned that 
the superconductivity is not destroyed up to 44 GPa where $T_{c}$ is still as high as 12 K. 
\cite{struz} The intrinsic $dT_{c}/dP\approx -1.1$ K/GPa under hydrostatic pressure 
conditions has recently been reported by other three groups.\cite{struz,schla,tang} We also 
have learned a possible explanation given by Tissen $et$ $al.$\cite{tissen} for the large 
$-dT_{c}/dP$ observed in the low $T_{c}=37.4\pm0.1$ samples.

\end{document}